# Giant Nernst Angle in Self-Intercalated van der Waals Magnet Cr$_{1.25}$Te$_2$

Shuvankar Gupta[1*], Olajumoke Oluwatobiloba Emmanuel[1*], Yasemin Ozbek[1], Mingyu Xu[2], Weiwei Xie[2], Pengpeng Zhang[1], and Xianglin Ke[1]

[1]Department of Physics and Astronomy, Michigan State University, East Lansing, Michigan 48824-2320, USA

[2]Department of Chemistry, Michigan State University, East Lansing, Michigan 48824, USA

*: These authors contributed equally to this work.

**Abstract:**

The discovery of two-dimensional van der Waals (vdW) magnetic materials has propelled advancements in technological devices. The Nernst effect, which generates a transverse electric voltage in the presence of a longitudinal thermal gradient, shows great promise for thermoelectric applications. In this work, we report the electronic and thermoelectric transport properties of Cr$_{1.25}$Te$_2$, a layered self-intercalated vdW material which exhibits an antiferromagnetic ordering at $T_N \sim$ 191 K followed by a ferromagnetic-like phase transition at $T_C \sim$171 K. We observe a prominent topological Hall effect and topological Nernst effect between $T_C$ and $T_N$, which is ascribable to non-coplanar spin textures inducing a real-space Berry phase due to competing ferromagnetic and antiferromagnetic interactions. Furthermore, we show that Cr$_{1.25}$Te$_2$ exhibits a substantial anomalous Nernst effect, featuring a giant Nernst angle of ~37% near $T_C$ and a maximum Nernst thermoelectric coefficient of 0.52 µV/K. These results surpass those of conventional ferromagnets and other two-dimensional vdW materials, highlighting Cr$_{1.25}$Te$_2$ as a promising candidate for advanced thermoelectric devices based on the Nernst effect.

**Introduction:**

The anomalous Nernst effect (ANE), driven by nontrivial topological Berry curvature in various magnets, has recently garnered increasing interest for its potential in thermoelectric applications [1–5]. Akin to the anomalous Hall effect (AHE) [6], the ANE produces a transverse thermoelectric voltage, orthogonal to the temperature gradient and magnetic field. However, unlike the AHE, which depends on the Berry curvature across the entire Fermi-sea, the ANE is particularly sensitive to the Berry curvature near the Fermi level. This makes it a key indicator of thermoelectric efficiency in materials with significant Berry curvature [5,7–10] and in certain ferromagnetic compounds [11,12]. Furthermore, from the perspective of technological applications, Nernst devices offer advantages over Seebeck devices due to their simpler transverse geometry. Given that the ANE is less extensively studied than the Seebeck effect, there is significant potential for advancing our understanding and control of the ANE through the synthesis of novel materials, which may pave the way for more efficient energy-harvesting technologies.

Since the discovery of intrinsic 2D magnetism in van der Waals (vdW) materials, layered magnets have emerged as attractive platforms for technological applications [13–15]. Of particular interest is the ability to systematically tune electron filling and magnetic properties in these materials without significantly altering their crystal structure. One of the recent research focuses is on transition metal dichalcogenides (TMDs) due to their flexibility in accommodating intercalated atoms between weakly coupled vdW atomic layers, enabling tuneable magnetic properties [16–21] and various physical properties [22–24]. Among these, $Cr_{(1+x)}Te_2$ stands out as a self-intercalated TMD with promising magnetic properties as demonstrated in various investigations [25–29]. Recent studies have reported giant AHE [30], magnetic skyrmions [31,32], tuneable Berry curvature [33], and room-temperature topological Hall effect (THE) in $Cr_{(1+x)}Te_2$ [34]. Additionally, it was reported that the Cr - Te system may exhibit Néel-type skyrmion features [35–37], however, it is noteworthy that some magnetization measurements [35–37] do not reveal any evidence of non-collinear magnetic moments, nor are such features observed in electronic transport studies [37]. Similarly, a recent investigation of the tr-$Cr_5Te_8$ (trigonal) system identified Néel-type skyrmions but found no indication of THE in the corresponding resistivity data [37]. In contrast, another study on the $Cr_{(1+x)}Te_2$ series of compounds reported a clear THE which was attributed to Néel-type skyrmions [32]. The controversy among these observations remains unresolved and warrants further investigation. Furthermore, while the AHE and THE have been relatively well-studied [30,34,38–40], the Nernst effect remains largely unexplored in vdW materials.

In this study, we present a comprehensive investigation of the electronic and thermoelectric transport properties of the $Cr_{1.25}Te_2$ compound. This material shows a giant Nernst angle ($S_{yx}/S_{xx}$, where $S_{yx}$ and $S_{xx}$ are the Nernst and Seebeck coefficient, respectively) of approximately 37%, which is comparable to the recently reported ferromagnetic compound $Fe_3GaTe_2$ (~ 62%) [41] and exceeds all other values reported previously [3,12,42–47] . The maximum Nernst coefficient is found to be 0.52 μV/K (at 160 K and 1 T), larger than the 0.3 μV/K observed for $Fe_3GeTe_2$ [42] and the 0.44 μV/K for $Fe_3GaTe_2$ [41]. Moreover, we report, for the first time, the observation of the topological Nernst effect in a van der Waals material series. The remarkable Nernst angle and Nernst responses make $Cr_{1.25}Te_2$ a highly promising candidate for the development of next-generation transverse thermoelectric devices.

**II. Method**

Single crystals of $Cr_{1.25}Te_2$ were synthesized using the self-flux method. High-purity chromium and tellurium with a molar ratio of 18:82 was placed in an alumina crucible which was then sealed within a vacuum-tight quartz tube. This assembly was then heated in a box furnace to 1000°C, maintained at this temperature for 10 hours, and subsequently cooled at a rate of 2 K/hour down to 800°C. The furnace was held at 800°C for 48 hours before the tube was removed from the furnace and centrifuged to eliminate excess flux. This process yielded large, shiny single crystals measuring approximately 2 mm × 2 mm × 1 mm. The compound composition of the crystals was determined using scanning electron microscopy equipped with energy-dispersive X-ray spectroscopy (EDX), utilizing a JEOL 7500F ultra-high-resolution instrument. The crystal structure was examined by X-ray diffraction (XRD) using Rigaku XtalLAB Synergy Dualflex Hypix single-crystal X-ray diffractometer. And surface topology was measured using scanning tunneling microscopy (STM). Detailed information about single crystal X-ray diffraction and STM measurements are described in the supplementary materials [48]. Magnetic susceptibility measurements were performed using a Superconducting Quantum Interference Device (SQUID) magnetometer and electronic transport measurements were conducted using a Physical Property Measurement System (PPMS) cryostat. A custom-designed sample puck compatible with the PPMS cryostat was employed for thermal and thermoelectric transport measurements. Temperature readings were acquired using type-E (Chromel-Constantan) thermocouples. A Keithley K2182A Nanovoltmeter was utilized to measure the thermoelectric voltage. The "cold end" of the sample was affixed to a high-conductivity, oxygen-free copper block using silver epoxy. A resistive heater (~1 kΩ) was attached to the opposite ("hot") end of the sample, and a heat current ($J_Q$) was applied within the *ab*-plane. The external magnetic field was oriented along the *c*-axis (out-of-plane direction). The $Cr_{1.25}Te_2$ sample demonstrates excellent air stability. All measurements were conducted on contacts made under ambient conditions. Over several months of study, no degradation or changes in structural, magnetic, or transport properties were observed.

**III. Results and Discussion**

Figure 1(a) and 1(b) present typical single-crystal XRD patterns and Tables S1

and S2 in the Supplementary Materials [48] show the refinement results. The structure was solved and refined using the Bruker SHELXTL Software Package [49,50] with the space group $P$-$3m$1 [48]. The composition was determined to be $Cr_{1.25(3)}Te_2$, which is consistent with the EDX analysis. Figure 1(c) displays the refined crystal structure of $Cr_{1.25}Te_2$ viewed along the a-axis. The red and blue spheres correspond to the Cr and Te atoms, respectively. Te and Cr1 layers alternate in the crystal structure, with Cr2 vacancies present in every other Cr layer. Note that while some extra peaks are observed, the intensity is too low to convincingly conclude that the Cr2 vacancies form long-range ordered superstructures (see details shown in Table S3 and the related discussion in the Supplementary Materials [48]. Figure 1(d) presents the STM image of an in-situ exfoliated $Cr_{1.25}Te_2$ crystal surface. The in-plane lattice constant was estimated to be 3.9 ± 0.1 Å, which is consistent with the expected values based on XRD measurements. The dark, circular features could be attributed to either Cr or Te vacancies in the lattice. The somewhat brighter and much larger hexagonal features that vary in size and orientation are thought to be electronic signatures from layers beneath the surface since STM convolutes geometric and electronic information.

To elucidate the magnetic properties of the material, we measured the temperature-dependent magnetization $M(T)$ with the magnetic field applied along two key crystallographic directions: the *c*-axis and the *ab*-plane. Figures 2(a) and 2(b) illustrate the $M(T)$ curves for $Cr_{1.25}Te_2$ in the zero-field-cooled (ZFC), field-cooled cooling (FCC) modes of $Cr_{1.25}Te_2$, under 0.1 T magnetic field. Along both crystallographic orientations, we observe distinct anomalies: a subtle kink appears at 191 K, identified as $T_N$, which corresponds to the antiferromagnetic (AFM) to paramagnetic (PM) transition; and an abrupt change around 171 K, marked as $T_C$, are indicative of a ferromagnetic-like (FM) to AFM transition. These transitions are further substantiated by the anomalies in the derivatives *dM/dT*, as shown in the insets of Figure 2(a) and 2(b).

Interestingly, previous studies of $Cr_{1+x}Te_2$ have consistently reported ferromagnetic ordering, with the transition temperature $T_C$ increasing as the concentration of excess Cr, denoted by *x*, increases [25,27,32]. This makes our observation of both ferromagnetic and antiferromagnetic orderings in $Cr_{1.25}Te_2$ particularly compelling, necessitating a reconsideration of the magnetic properties in this system. We speculate that the intercalated Cr2 atoms play a decisive role in determining the nature of magnetic ordering. Notably, the existence of both FM and AFM phases has been reported in literature for $Cr_2Te_3$ [51,52], though recent theoretical studies suggest a more nuanced magnetic phase [51–54]. These works on $Cr_2Te_3$ [52–54] argue that, rather than simple ferromagnetic ordering with moments aligned along the *c*-axis, the system favours a canted antiferromagnetic configuration [52–54]. In this scenario, magnetic moments possess large out-of-plane components oriented parallel to the *c*-axis, while smaller in-plane components are arranged antiparallel in the *ab*-plane. It was further proposed that the Cr atoms intercalated between the $CrTe_2$ layers experience a stronger canting of their magnetic moments compared to those within the $CrTe_2$ layers [52–54]. This complex magnetic structure implies delicate competition between FM and AFM interactions in the $Cr_{1+x}Te_2$ system, which accounts for the observation of both FM and AFM transitions in our $Cr_{1.25}Te_2$ compound.

The magnetic anisotropy of the system is affirmed by the isothermal magnetization measurements conducted at 5 K, as shown in Figure 2(c). Here, magnetization along the *c*-axis saturates more readily compared to that measured along the *ab*-plane, firmly establishing the *c*-axis as the easy-axis of magnetization. This result is in agreement with previous reports on the $Cr_{1+x}Te_2$ series [25,27,32]. The substantial difference in magnetization between the two directions indicates pronounced uniaxial magneto crystalline anisotropy in $Cr_{1.25}Te_2$, which aligns with recent studies highlighting exceptionally large anisotropy in ferromagnetic $Cr_2Te_3$ [54].

Isothermal magnetization measurements at various temperatures between $T_C$ and $T_N$, shown in Figure 2(d), offer additional insight into the system's magnetic behaviour. A clear metamagnetic transition is observed at $T_C < T < T_N$ (e.g., 175 K, 180 K, and 185 K). This metamagnetic transition corresponds to the field-driven evolution of the magnetic state from AFM state to polarized FM state. Such behaviour affirms the delicate balance between FM and AFM interactions in this material and highlights the rich and tuneable magnetic landscape that exists in this family of compounds. Our study thus presents a comprehensive picture of the magnetism in $Cr_{1.25}Te_2$, bridging experimental observations with theoretical predictions [52–54]. To provide a more complete description of the magnetic behaviour, we have measured the magnetic field dependence of magnetization for H ⊥ c at various temperatures (Figure S1 [48]). The results indicate that metamagnetic transitions observed for H ∥ c is absent in the H ⊥ c configuration, affirming the strong uniaxial magnetic anisotropy discussed above. The implication of magnetic anisotropy on electronic and thermoelectric transport properties is discussed in the Supplementary Information [48].

The temperature dependence of resistivity ($\rho_{xx}$) and longitudinal Seebeck coefficient ($S_{xx}$) is depicted in Figures 3(a). Both exhibit a marked anomaly near $T_C$, indicating strong magnetoelectric coupling. As $T$ approaches $T_C$ upon warming, thermal agitation disrupts the magnetic moment alignment, enhancing charge carrier scattering, which in turn increases resistivity and modifies the Seebeck coefficient. Critical magnetic fluctuations and intensified spin-dependent scattering in this region further amplify these effects. Notably, electron-magnon scattering near $T_C$ plays a pivotal role in modulating both electrical and thermoelectric properties, leading to pronounced changes in the slopes of ($\rho_{xx}$) and ($S_{xx}$). These observations underscore the significant impacts of spin ordering and magnetic fluctuations on transport phenomena, particularly around the ferromagnetic phase transition.

The $Cr_{1.25}Te_2$ compound also exhibits pronounced features in magnetic field dependence of magnetoresistance (MR) and the magneto-Seebeck (MS) coefficient as shown in Figure 3(b) and 3(c), respectively, which are associated with its magnetic ordering transitions. Below $T_C$ (171 K), the ferromagnetic-like ordering leads to a low MR value (~1-2%) as well as low MS value (~5%) with a nearly linear field dependence of both MR and MS. In the temperature range between $T_C$ and $T_N$, both MR and MS undergo a change in slope around the spin-flop phase transition field observed in $M(H)$ curves shown in Fig. 2(d). Figure 3(d) summarizes the temperature dependence of the extracted values of MR and MS measured at 1 T, from which large changes in both MR and MS values are only visible in the temperature regime between $T_C$ and $T_N$. Specifically, near $T_C$ both MR and MS reach a maximum around -10% and +27%, respectively. These features are consistent with the temperature dependence of $\rho_{xx}$ and $S_{xx}$ shown in Fig. 3(a) which indicates that FM spin alignment gives rise to smaller $\rho_{xx}$ and larger $S_{xx}$.

The MR and MS measurements display distinct trends across the investigated temperature range. In the FM region, both MR and MS exhibit a nearly linear dependence on the applied magnetic field, which is attributed to the domain alignment and minimal spin scattering. In contrast, in the antiferromagnetic phase region ($T_C < T < T_N$), the MR and MS show a large change in slopes when approaching the spin-flop phase transition.

For MR, the slope changes can be explained by the evolution of magnetic structure and the resulting spin scattering effect as a function of magnetic field. Upon approaching the spin-flop transition, the spin scattering reduces which causes a large drop in MR. Above the spin-flop transition, the slope of MR decreases due to the relatively smaller change in spin scattering effect when the spins gradually tilt from canted antiferromagnetic state to the fully polarized ferromagnetic state. The MS effect is particularly sensitive to spin-polarized thermoelectric transport, which depends on the asymmetry of the electronic density of states (DOS) near the Fermi level and the spin configuration. In the metamagnetic region, the evolving magnetic structure modifies the DOS asymmetry and scattering rates, resulting in the observed slope changes.

Additionally, an anti-correlation between MR (negative) and MS (positive) is observed in the metamagnetic transition region, with MR be negative and MS positive. This behaviour can be understood through the Mott relation, $S = -\frac{\pi^2 k_B^2 T}{3e\sigma} \frac{d\sigma}{d\zeta}|_{E_F}$ ($E_F$ is the Fermi energy). If $\frac{d\sigma}{d\zeta}|_{E_F}$ is nearly field-independent, the MS is proportional to the resistivity $\rho$ ($=1/\sigma$), which is evident in the FM region where both MR and MS are negative. However, in the transition region, the metamagnetic transitions induce changes in the electronic structure, making $\frac{d\sigma}{d\zeta}|_{E_F}$ field-dependent and possibly exhibiting an opposite trend. This can dominate the magnetic field dependence of resistivity, resulting in the observed anti-correlation between MR and MS. Such changes in $\frac{d\sigma}{d\zeta}|_{E_F}$ may be linked to alterations in the DOS or carrier dynamics upon the spin-flop phase transition.

We now discuss the transverse electronic and thermoelectric response in $Cr_{1.25}Te_2$. Figure 4(a) shows the transverse resistivity ($\rho_{yx}$) versus. magnetic field along the *c*-axis, with the current applied along the x-axis and voltage measured along the y-axis. The total Hall resistivity ($\rho_H$) is expressed as $\rho_H = \rho_N + \rho_A + \rho_T$, where $\rho_N$, $\rho_A$, and $\rho_T$ represent normal, anomalous, and topological contributions, respectively. In the high-field region where $\rho_T$ vanishes, $\rho_H$ simplifies to $R_0 H + R_S 4\pi M$. The slope $R_0$ and intercept $4\pi R_S$ are estimated by analysing the linear $\rho_H/M$ versus $H/M$ relation. As an example, following this method we extracted the contributions of $\rho_N$, $\rho_A$, and $\rho_T$ at 175 K, as shown in Figure 4(b). Figure 4(c) presents $\rho_T$ as a function of the magnetic field, extracted at various temperatures for the regime $T_C < T < T_N$. Figure 4(d) represents the field dependence of Hall conductivity ($\sigma_{xy}$) taken at different temperatures below $T_C$.

The THE observed in $Cr_{1.25}Te_2$ can be attributed to the presence of non-coplanar spin structures, such as skyrmions or chiral spin textures [55–57]. At $T_C < T < T_N$, the interplay among ferromagnetic and antiferromagnetic exchange interactions as well as Zeeman interaction may lead to non-coplanar spin textures. Such structures can generate a real-space Berry phase that acts on conduction electrons, leading to an additional transverse voltage distinct from the conventional anomalous Hall effect. It is worth noting that future neutron diffraction studies of $Cr_{1.25}Te_2$ would be valuable in elucidating the magnetic structure and resolving the underlying cause of the controversial THE effect observed in the $Cr_{1+x}Te_2$ series [35,37] as discussed above. This makes $Cr_{1.25}Te_2$ an exceptional compound for understanding the nature of magnetism in the $Cr_{1+x}Te_2$ series.

Figure 5(a) presents the magnetic field dependence of the transverse thermoelectric response ($S_{yx}$) of $Cr_{1.25}Te_2$ at various temperatures, closely resembling the Hall data shown in Figure 4(a). This similarity suggests shared mechanisms behind both Hall and Nernst effects, driven by non-trivial band topology and finite spin chirality. Similar to the Hall effect, $S_{yx}$ can be decomposed as $S_{yx} = S^0_{yx} + S^A_{yx} + S^T_{yx}$, representing the normal, anomalous, and topological contributions, respectively. Using the above methodology, these three Nernst components were extracted, as shown in Figure S2 [48] for the data measured at 173.6 K. The field dependence of $S^T_{yx}$ at various temperatures within the region ($T_C < T < T_N$) is illustrated in Figure S3 [48]. To further investigate the thermoelectric properties, we estimated the temperature dependence of Nernst thermoelectric conductivity ($\alpha_{xy}$) as shown in Figure 5(b) using the measured resistivities ($\rho_{xx}$ and $\rho_{yx}$), and thermopowers ($S_{xx}$, $S_{yx}$), following the relation: $\alpha_{xy} = S_{xy}\sigma_{xx} +$ $S_{xx}\sigma_{xy}$ [58]. The peak value of $\alpha_{xy}$ is 0.43 A/(m·K) at 50 K, which surpasses those of conventional ferromagnets [59–61] as well as other reported 2D vdW materials [41,42,62].

Figure 5(c) shows the temperature dependence of $S_{yx}$ (left axis), which increases with temperature and reaches a maximum value of 0.52 µV K$^{-1}$ at 151 K, close to $T_C$. This value exceeds those recently reported for other two-dimensional vdW materials, including 0.3 µV K$^{-1}$ for $Fe_3GeTe_2$ [42] and $Fe_{3-x}GeTe_2$ [62], and 0.44 µV K$^{-1}$ for $Fe_3GaTe_2$ [41]. A recent study of the $Cr_{(1+x)}Te_2$ [33] systems indicates that the Berry curvature in these magnetic semimetals is highly tuneable. For instance, it has been shown that the AHE in $Cr_{(1+x)}Te_2$ thin films can flip its sign across different doping levels, a phenomenon linked to changes in the Berry curvature as the Fermi energy shifts [33]. As discussed earlier, the ANE is more sensitive to the Berry curvature of electronic bands near the Fermi level, making it a more effective probe of the electronic structure. Thus, the significant ANE observed in $Cr_{1.25}Te_2$ suggests its topological electronic band structure with a large Berry curvature near the Fermi level. To provide a comprehensive evaluation of $Cr_{1.25}Te_2$ thermoelectric potential, we calculated the thermoelectric figure of merit (ZT) (Figure S4 [48]) for the Nernst configuration. The estimated ZT value is $4.5\times10^{-6}$ at $T_C$. Note that ZT for the Nernst configuration is generally much smaller compared to the ZT value for conventional thermoelectric devices based on Seebeck, since the Nernst coefficient is often 2-3 orders in magnitude smaller than Seebeck coefficient.

To further shed light on the observed Nernst effect of $Cr_{1.25}Te_2$, we plotted Nernst angle (=$S_{yx}/S_{xx}$) (right Figure 5(c)) as a function of temperature. As shown, a peak with magnitude ~37 % occurs near $T_C$. This giant value of anomalous Nernst angle is comparable to the recently reported for $Fe_3GaTe_2$ [41] and surpasses all other previous values reported in the literature [42,44,46,63–68]. In Figure 5(d), we made a comparison of the anomalous Nernst angle with existing systems in literature [42,44,46,63–68]. The giant response of Nernst angle of $Cr_{1.25}Te_2$ stems from the large value of $S_{yx}$ and small value of $S_{xx}$ (Figure 3(a)), the latter of which may result from the opposite moving directions of up-spin and down-spin charge carriers in the presence of thermal gradient, a mechanism that was recently proposed for $Fe_3GaTe_2$ [41]. The observation of giant Nernst angle along with large Nernst coefficients compared to the existing vdW materials such as $Fe_3GeTe_2$ [42], $Fe_{3-x}GeTe_2$ [62] and $Fe_3GaTe_2$ [41] makes $Cr_{1.25}Te_2$ promising materials for practical applications for Nernst effect-based device.

## IV. Conclusion

In summary, $Cr_{1.25}Te_2$, which undergoes an antiferromagnetic phase transition followed by ferromagnetic-like ordering upon cooling, exhibits exceptional transverse transport properties underpinned by non-trivial band topology and intricate spin textures. This material exhibits both a topological Hall effect and a topological Nernst effect in the temperature regime between $T_C$ and $T_N$, which are associated with a non-coplanar spin structure that generates a real-space Berry phase. Notably, $Cr_{1.25}Te_2$ showcases a remarkable anomalous Nernst effect, featuring a giant Nernst angle of approximately 37% near $T_C$. This performance surpasses that of conventional ferromagnets, topological magnets, and other two-dimensional van der Waals materials. The combination of this giant Nernst angle and Nernst responses positions $Cr_{1.25}Te_2$ as a highly promising candidate for next-generation transverse thermoelectric devices.

## ACKNOWLEDGMENTS

S. G., O.E., and X.K. acknowledge the financial support by the U.S. Department of Energy, Office of Science, Office of Basic Energy Sciences, Materials Sciences and Engineering Division under Grant No. DE-SC0019259. The electronic and thermoelectric transport measurements were supported by National Science Foundation (DMR-2219046). STM experiments were supported by the U.S. Department of Energy (DOE), Office of Basic Energy

Sciences, Division of Materials Sciences and Engineering under Award Number DE-SC0019120. P. P. Zhang acknowledges the financial support from National Science Foundation (DMR-2112691). M.X. and W.X. acknowledges the financial supported by the U.S.DOE-BES under Contract DE-SC0023648.

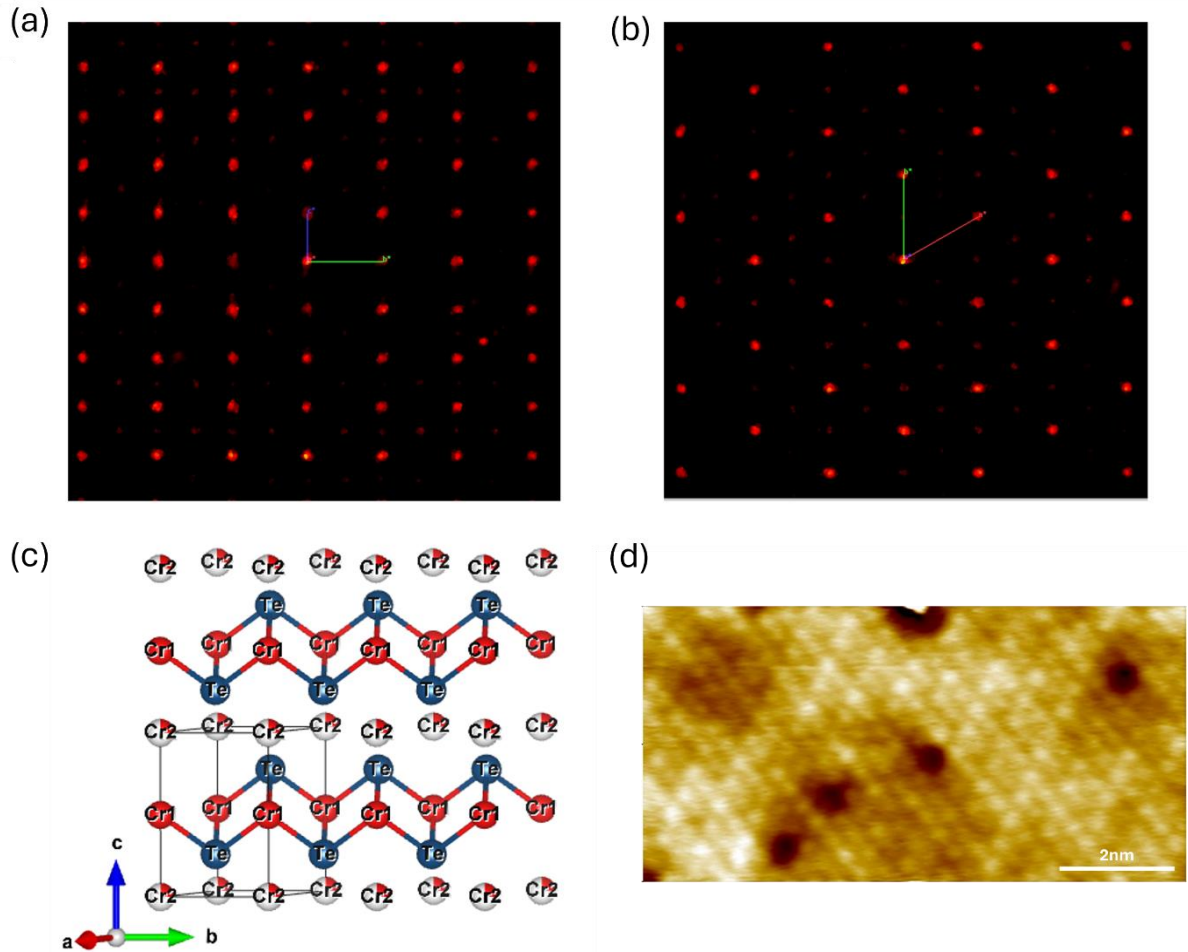

Figure 1: Single crystal X-ray reciprocal map of (a) (0kl) and (b) (hk0) of $Cr_{1.25(3)}Te_2$. (c) Crystal structure of $Cr_{1.25}Te_2$. (d) A close-up STM image ($V_s = -0.05$ V, $I_t = 600$ pA, where $V_s$ is the voltage/bias applied to the sample (the 0 V point being the tip) and $I_t$ is the tunnelling current used between the tip and the sample.).

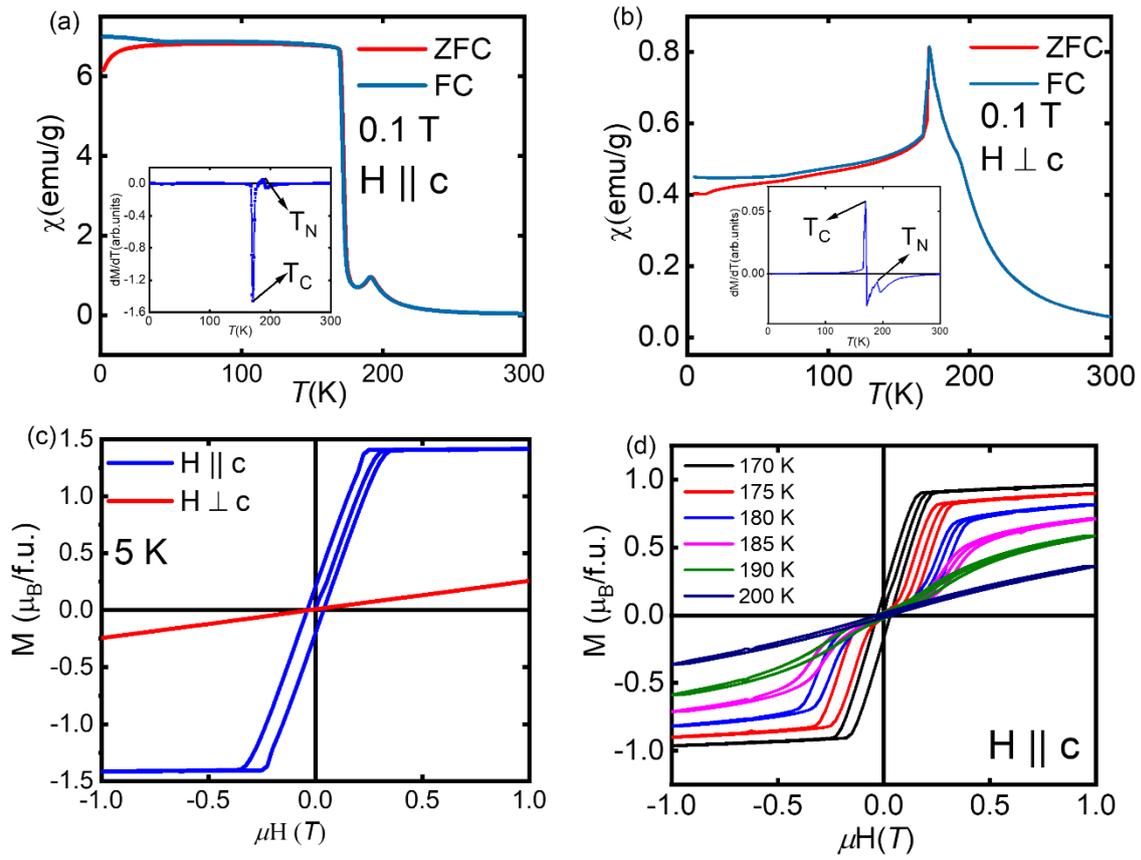

Figure 2: (a) Temperature dependence of magnetic susceptibility of $H \parallel c$ and its derivative (inset). (b) Temperature dependence of magnetic susceptibility of $H \perp c$ and its derivative (inset). (c) Magnetic field dependence of magnetization measured at 5K for both $H \parallel c$ and $H \perp c$. (d) Magnetic field dependence of magnetization $H \parallel c$ measured at various temperatures close to $T_C$ and $T_N$.

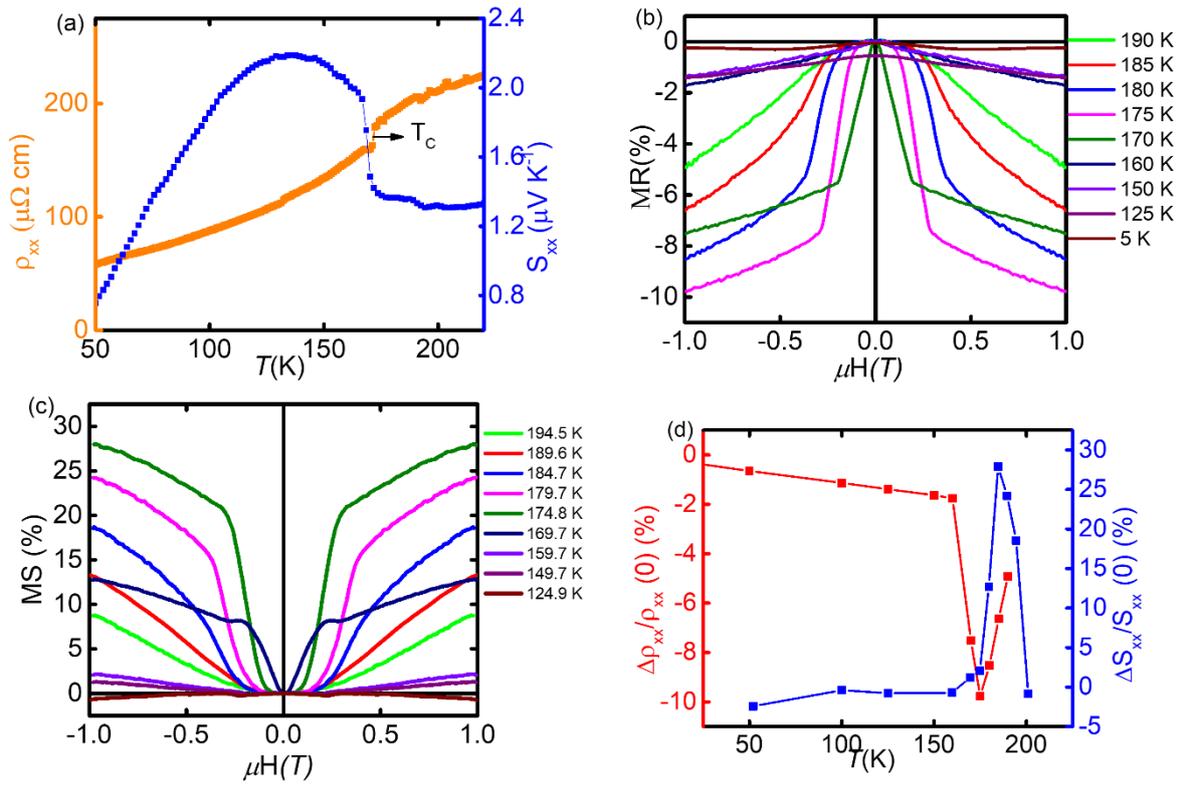

Figure 3: (a) Temperature dependence of resistivity and Seebeck. (b) Field dependence of MR at various temperature. (c) Magnetic field dependence of magneto Seebeck coefficient at various temperature. (d) Temperature dependence of MR and MS measured at 1 T. The field was applied parallel to the c-axis (H∥c).

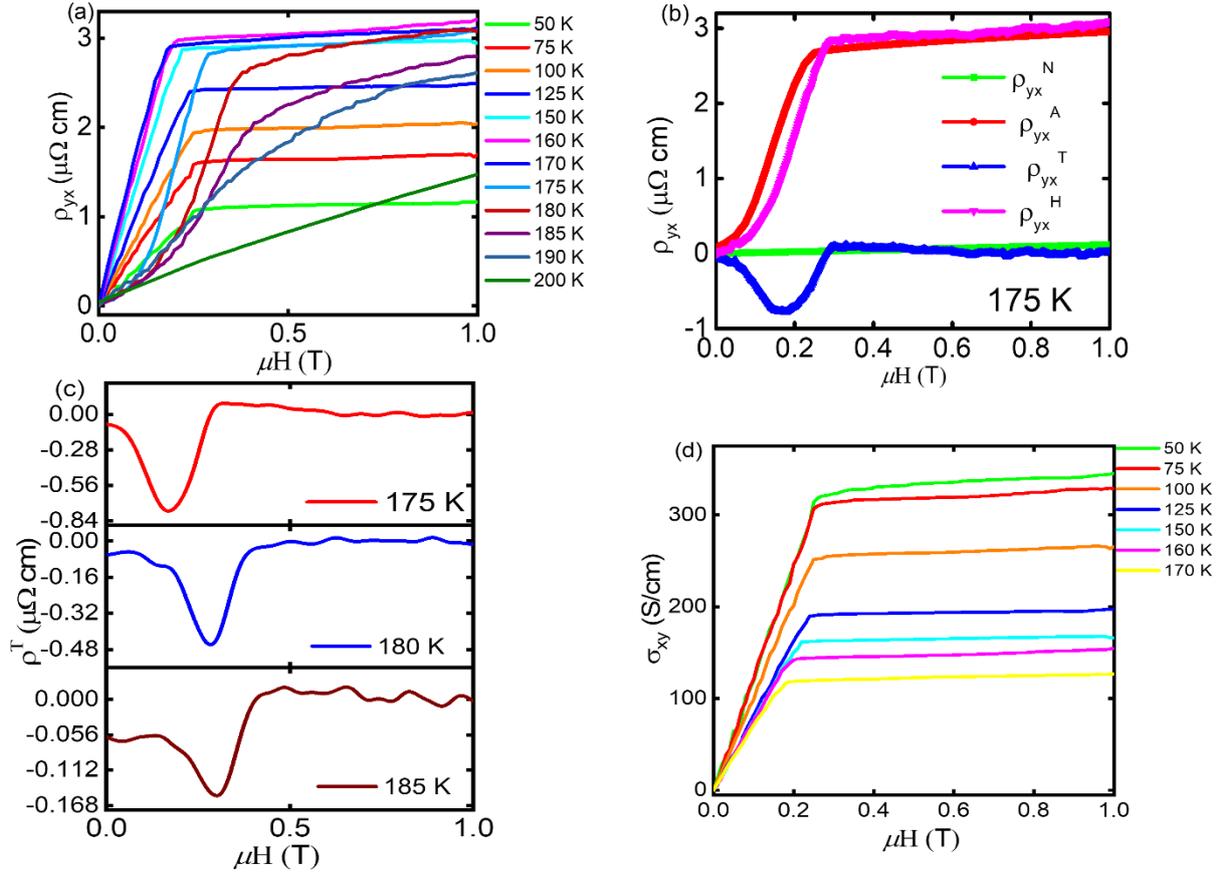

Figure 4: (a) Magnetic field dependence of Hall resistivity measured at different temperatures. (b) Magnetic field dependence of different Hall contributions measured at 175 K. (c) Field dependence of the extracted topological Hall resistivity at various temperatures. (d) The magnetic field dependence of Hall conductivity measured at various temperatures. The field was applied parallel to the c-axis (H∥c).

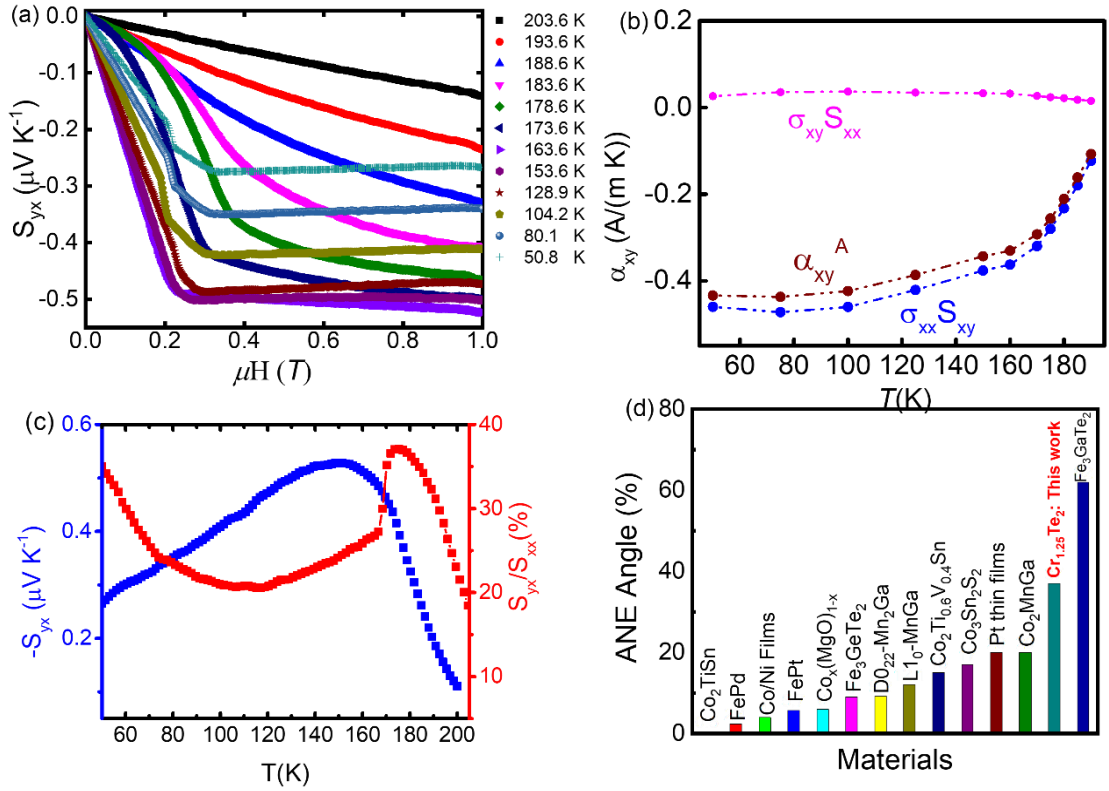

Figure 5: (a) Magnetic field dependence of Nernst coefficients measured at various temperatures. The field was applied parallel to the c-axis (H∥c). (b) Temperature dependence of the calculated Nernst thermoelectric conductivity. (c) Temperature dependence of Nernst coefficient and Nernst angle. (d) Comparison of Nernst angle of various state-of-the-art materials reported in literature [38,39,42,62–67] with $Cr_{1.25}Te_2$.